\documentclass[RNAAS]{aastex62x}

\begin{document}

\title{The Intermediate Polar FO~Aquarii Has Not Been the Same Since Recovering from a Series of Low States}

%% Note that the corresponding author command and emails has to come
%% before everything else. Also place all the emails in the \email
%% command instead of using multiple \email calls.
\correspondingauthor{Peter Garnavich}
\email{pgarnavi@nd.edu}

\author{Peter Garnavich}
\affiliation{Department of Physics and Astronomy, University of Notre Dame, Notre Dame, IN 46556, USA}

\author{Colin Littlefield}
\affiliation{Bay Area Environmental Research Institute, Moffett Field, CA 94035 USA}
\affiliation{Department of Physics and Astronomy, University of Notre Dame, Notre Dame, IN 46556, USA}

\author{Rebecca S. Boyle}
\affiliation{Department of Physics and Astronomy, University of Notre Dame, Notre Dame, IN 46556, USA}

\author{Mark Kennedy}
\affiliation{University College Cork, Cork, Ireland}

\begin{abstract}

FO~Aqr is a bright intermediate polar that has long displayed large amplitude photometric variations corresponding to the 20.9 min spin period of its white dwarf. Between 2016 and 2020, the system suffered a series of unprecedented low-states, but recent data shows that it has now recovered to its normal optical luminosity. We compare the light curves obtained by K2/Kepler in 2014 with photometry from the TESS mission obtained in 2021. We find that the spin pulse that had been the dominant feature of the light curve in 2014 is now weak over the second half the binary orbit and that a beat pulse is enhanced in the TESS photometry. Variations at approximately twice the spin frequency are now seen over the second half of the orbit. These photometric properties may be the new normal for FO~Aqr now that its white dwarf has begun to spin down.

\end{abstract}

%% See the online documentation for the full list of available subject
%% keywords and the rules for their use.

%% Start the main body of the article. If no sections in the 
%% research note leave the \section call blank to make the title.
\section{Introduction} 

%%%%
\renewcommand{\thefigure}{1a}
\begin{figure}
\begin{center}
\includegraphics[scale=0.46,angle=0]{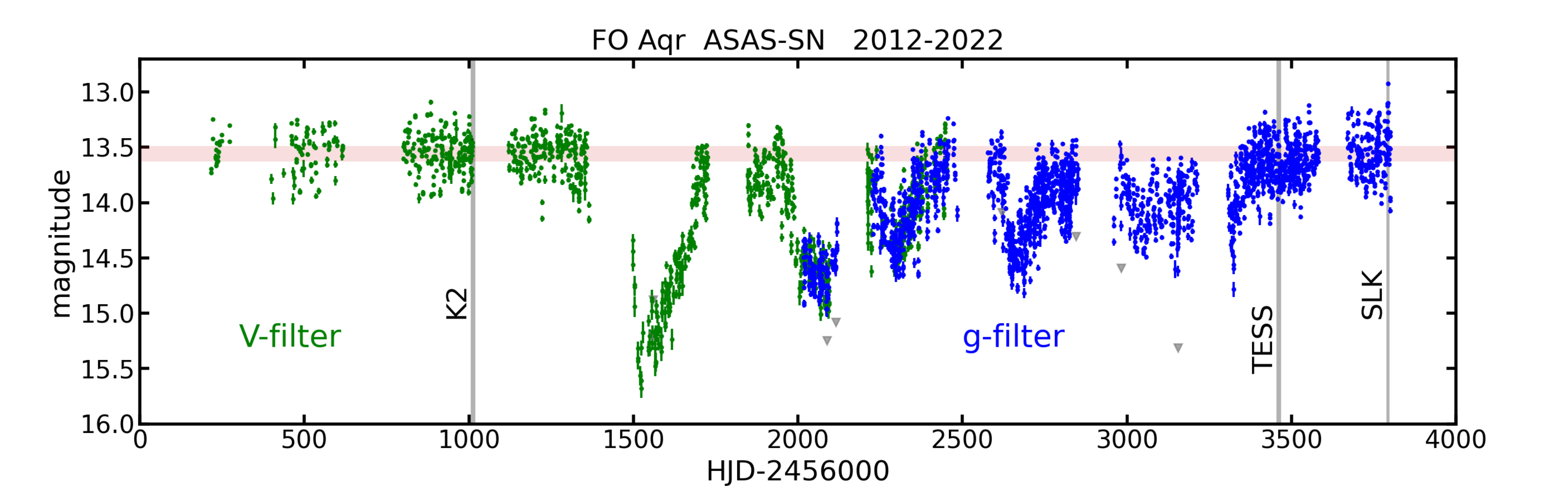}
\caption{The full ASAS-SN light curve of FO~Aqr covering 2012 to 2022. The red band shows the average brightness of FO~Aqr excluding the low states. Vertical lines mark the dates of the K2, TESS, and SLK observations.  \label{asassn}}
\end{center}
\end{figure}
%%%%

FO~Aquarii (FO~Aqr) is a bright Intermedate Polar (IP) that has been dubbed the "King of IPs" \citep{patterson83} because of its regular, large amplitude optical pulses seen at the spin period of its white dwarf (WD) primary. IPs are cataclysmic variables (CVs) consisting of a magnetic WD primary and a mass donating secondary star. In FO~Aqr, the orbital period is 4.85~hr, while the WD rotates every 20.9~min. The strong spin pulse in the light curve of FO~Aqr has allowed precise measurement of the variations in the WD spin period \citep{patterson20,littlefield20}.

For decades, FO~Aqr was observed at a fairly steady average brightness of $V\approx 13.5$~mag \citep[e.g.][]{garnavich88}. However, in 2016 the system began an unprecedented series of low-states, sometimes fading by more than two magnitudes \citep{littlefield16}. As shown by the ASAS-SN light curve \citep{kochanek17} in Figure~1a, the depth of these sputtering low states have declined over the years and the system appears to have recovered to its ``normal'' optical brightness in 2021.

%%%%
\renewcommand{\thefigure}{1b}
\begin{figure}
\begin{center}
\includegraphics[scale=0.35,angle=0]{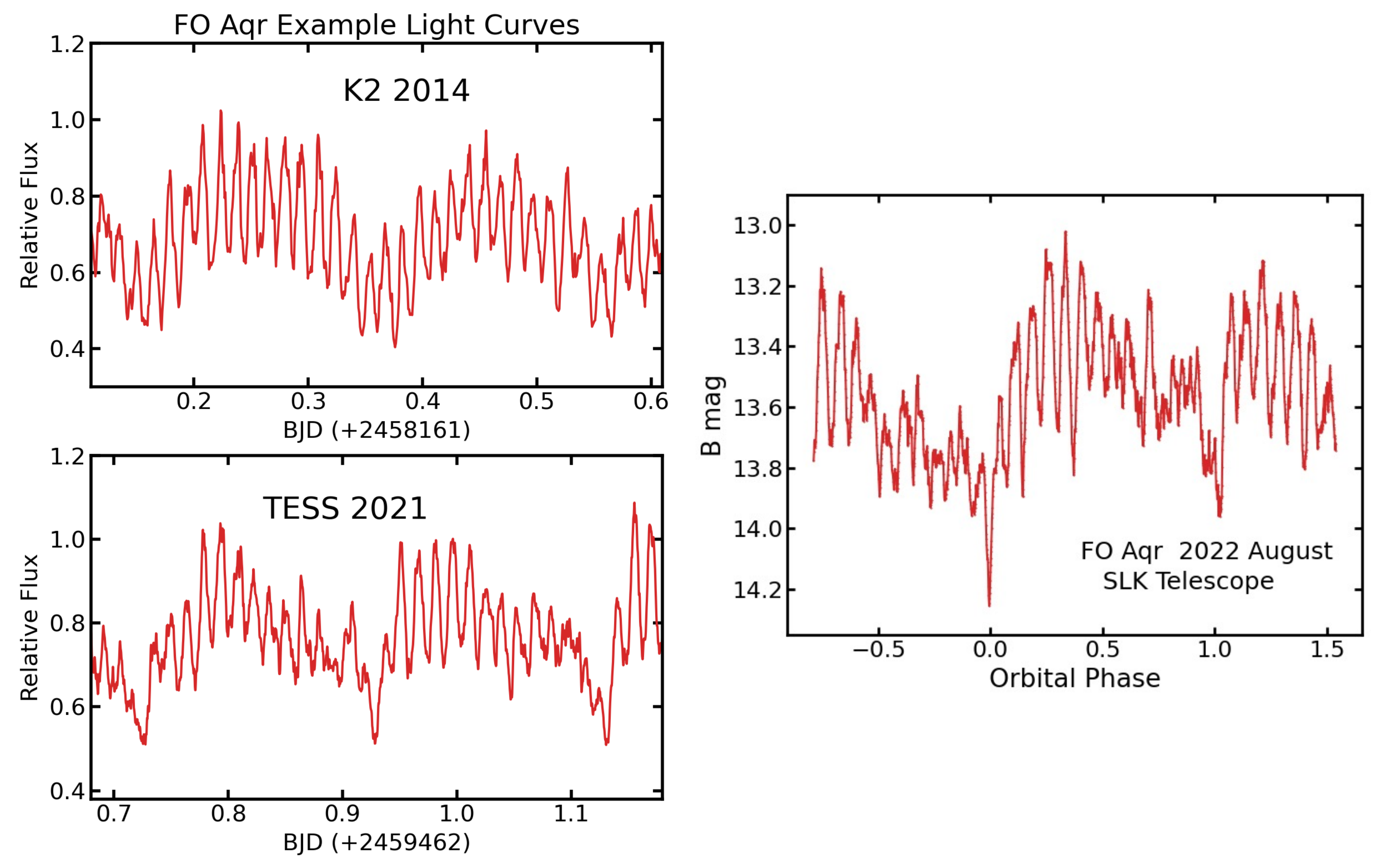}
\caption{Light curves of FO~Aqr from 2014 (K2, {\bf upper-left}) and 2021 (TESS, {\bf lower-left}). Each plot displays two consecutive binary orbits with each minimum corresponding to partial eclipse and inferior conjunction of the secondary star. $B$-band light curves from two nights in 2022 August are shown on the right. \label{sample}}
\end{center}
\end{figure}
%%%%

%\vspace{0.8cm}
\section{Data}

FO~Aqr was observed by K2 during Campaign~3 between November 2014, and February 2015 with a 60~s cadence. \citet{kennedy16} provides an extensive analysis of these data. TESS observed FO~Aqr during sector~42 over August and September, 2021, with the 20~s cadence. 

The K2 and TESS data were queried using the Lightkurve python package \citep{lightkurve18} and SAP photometry was extracted using the default aperture. Typical light curves over two binary orbits are shown Figure~1b with the minima corresponding to partial eclipses marking inferior conjunction of the secondary. The 2014 K2 light curve shows strong spin pulses at all orbital phases. The 2021 TESS light curve displays weaker spin pulses over the second half of the orbit. Our multi-filter ground-based photometry suggests that bandpass differences between K2 and TESS are not to blame for the observed change in light-curve characteristics.

Ground-based photometry was obtained in 2022 August with the 0.8-m aperture Sarah L. Krizmanich (SLK) telescope. Twelve hours of data were obtained through $B$ and $I$ filters with 4.0~s exposures covering two full binary orbits (Figure~1b). Photometric zeropoints were estimated from the APASS calibration \citep{APASS}. 

We downloaded available photometry of FO~Aqr from the ASAS-SN Sky Patrol website and the light curve is plotted in Figure~1a. Before the onset of the low-states in 2016, the average brightness of FO~Aqr was $V=13.55\pm 0.01$~mag. After 2021 July, the average brightness was $g=13.59\pm 0.01$~mag. Between the K2 and TESS visits, ASAS-SN transitioned from observing in a $V$-band filter to SDSS $g$-band. Measurements in both filters were obtained in 2018, and we used this to estimate an average color for FO~Aqr of $g-V=0.05\pm 0.03$~mag. Thus, after correcting for the color of FO~Aqr, the average difference in optical luminosity before 2016 and after 2020 is only 0.01$\pm 0.03$~mag.

%%%%
\renewcommand{\thefigure}{1c}
\begin{figure}
\begin{center}
\includegraphics[scale=0.45,angle=0]{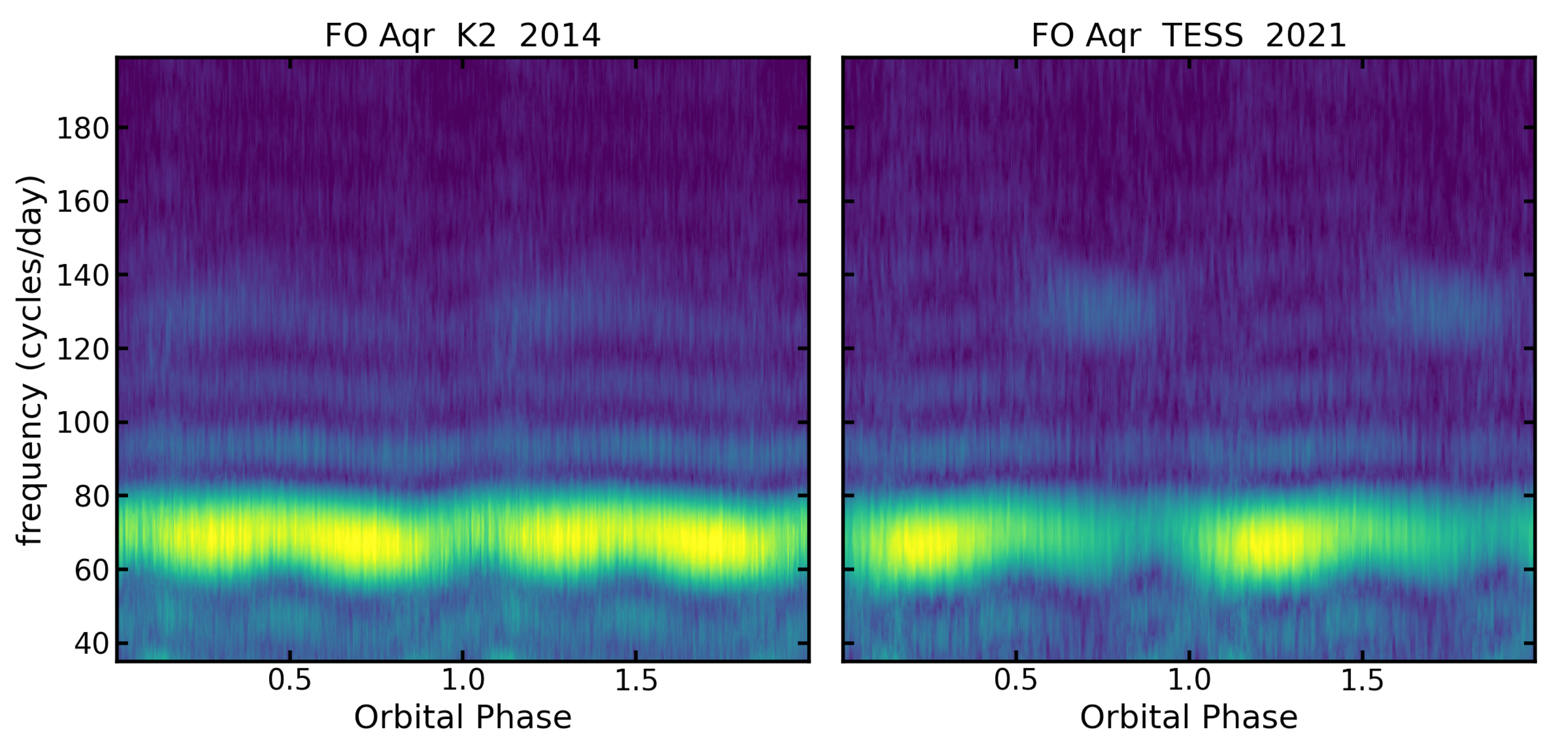}
\caption{The dynamic power spectrum of the K2 photometry in 2014 and TESS photometry in 2021. A signal at approximately twice the spin frequency appears between orbital phases 0.6 and 0.9 where the spin pulse appears weaker in 2021.  \label{power}}
\end{center}
\end{figure}
%%%%

\vspace{0.7 cm}
\section{Analysis \& Conclusion}

The example light curves shown in Figure~1b hint at a change in the photometric propterties of FO~Aqr between 2014 and 2021. To quantify this, we calculated dynamic Lomb-Scargle periodograms (L-S) \citep{lomb76,scargle82} in moving windows over the K2 and TESS data sets. To resolve the variations within an orbit, the width of the window was chosen to be 1.4~hr and it was stepped through the data in 15~min increments. The periodograms were then sorted on orbital phase and the results are shown in Figure~1c. The strongest signal is at a frequency of 68~c/d which corresponds to the WD spin ($\omega$). In the 2014 K2 data, the amplitude of the spin pulse is nearly equal between the first and second halves of the binary orbit. For the 2021 TESS photometry, it is clear that the spin pulse between orbital phases 0.5 and 1.0 is significantly weaker than between phases 0.0 and 0.5. There is also a signal detected at 130~c/d that is present only between orbital phases 0.6 and 0.9. This transient signal peaks close to twice the WD spin frequency ($2\omega$).  

We have created 2-dimensional phased light curves for the K2 and TESS data. These are images where the flux is phased on both the orbital and spin periods of FO~Aqr and these are displayed at Figure~1d. In addition to the suppressed pulse amplitude over the second half orbit, a diagonal band is detected running through orbital phases 0.5 to 0.9 that corresponds to a beat pulse $(\omega-\Omega$, where $\Omega$ represents the orbital frequency). This feature is in-phase and adds to the spin pulse over the first half of the orbit but becomes out of phase in the second half resulting in multiple, low-amplitude peaks with power near $2\omega$ \citep{littlefield16}. 

%%%%
\renewcommand{\thefigure}{1d}
\begin{figure}
\begin{center}
\includegraphics[scale=0.45,angle=0]{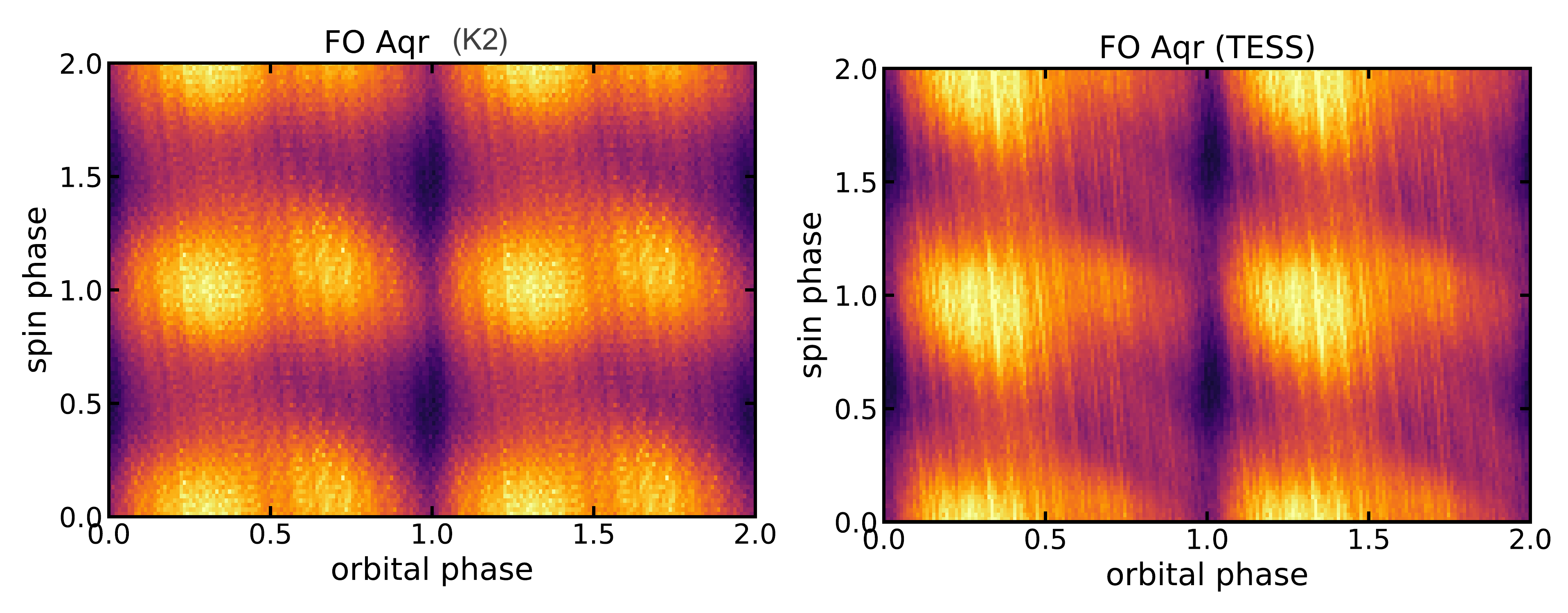}
\caption{The flux from FO~Aqr as a function of orbital phase and spin phase. For the TESS data, the diagonal feature seen around spin phase of 0.5 is the beat pulse that is not detectable in the K2 photometry.  \label{2d}}
\end{center}
\end{figure}
%%%%

FO~Aqr has appeared to recover from a series of low-states to reach a luminosity very close to its historical average. However, periodic variations in the light curve of FO~Aqr has not returned to its pre-2016 state. When compared with K2 obtained in 2014, the light curve over 2021/22 shows an enhanced beat pulse and a severe suppression of the spin pulse over the second half of the orbit. Before the series of low states, the FO~Aqr light curve displayed a strong spin pulse that dominated over all orbital phases. 

High cadence ground-based photometry in 2022 confirms the suppression of strong pulses in the second half of the orbit as seen in the 2021 TESS data. Short time-scale, irregular variability is now apparent in the later orbital phases.

The light curve properties of FO~Aqr may continue to evolve and eventually return to a spin pulse dominated state. Or, this may be the new normal for FO~Aqr light curve given the recent switch in sign of the WD spin derivative \citep{patterson20,littlefield20}.

\acknowledgments

We thank ASAS-SN for allowing public use of their data and creating an efficient interface to access it. We thank the Krizmanich family for their generous donation in funding the Sarah L. Krizmanich telescope. We acknowledge J. Crass for his help in obtaining the SLK telescope data.

%\begin{figure}
%\begin{center}
%\includegraphics[scale=0.46,angle=0]{rn_fig.pdf}
%\caption{{\bf Left Column:} Light curves of FO~Aqr from 2014 (K2, {\bf top}) and 2021 (TESS, {\bf middle}). Each plot displays two consecutive binary orbits with each minimum corresponding to partial eclipse and inferior conjunction of the secondary star. Light curves from two nights in 2022 August are shown at the {\bf bottom} of the left column.  {\bf Right Column:} The full ASAS-SN light curve of FO~Aqr covering 2012 to 2022 ({\bf top}). The red band shows the average brightness of FO~Aqr excluding the low states. {\bf middle:} The dynamic power spectrum of the K2 photometry in 2014 and TESS photometry in 2021. A signal at approximately twice the spin frequency appears between orbital phases 0.6 and 0.9. {\bf bottom:} The flux from FO~Aqr as a function of orbital phase and spin phase. For the TESS data, the diagonal feature seen around spin phase of 0.5 is the beat pulse that is not detectable in the K2 photometry.  \label{fig1}}
%\end{center}
%\end{figure}

\end{document}